%% file: verifiable-formal-requirements.tex
\documentclass{article}

\usepackage[left=3cm, right=3cm, top=2cm, bottom = 2cm]{geometry}

\usepackage[utf8]{inputenc}
\usepackage[T1]{fontenc}
\usepackage{lmodern}
\usepackage[UKenglish]{isodate}

\usepackage{scalerel}
\usepackage{tikz}
\usetikzlibrary{svg.path}

\definecolor{orcidlogocol}{HTML}{A6CE39}
\tikzset{
  orcidlogo/.pic={
    \fill[orcidlogocol] svg{M256,128c0,70.7-57.3,128-128,128C57.3,256,0,198.7,0,128C0,57.3,57.3,0,128,0C198.7,0,256,57.3,256,128z};
    \fill[white] svg{M86.3,186.2H70.9V79.1h15.4v48.4V186.2z}
                 svg{M108.9,79.1h41.6c39.6,0,57,28.3,57,53.6c0,27.5-21.5,53.6-56.8,53.6h-41.8V79.1z M124.3,172.4h24.5c34.9,0,42.9-26.5,42.9-39.7c0-21.5-13.7-39.7-43.7-39.7h-23.7V172.4z}
                 svg{M88.7,56.8c0,5.5-4.5,10.1-10.1,10.1c-5.6,0-10.1-4.6-10.1-10.1c0-5.6,4.5-10.1,10.1-10.1C84.2,46.7,88.7,51.3,88.7,56.8z};
  }
}

\newcommand\orcidicon[1]{\href{https://orcid.org/#1}{\mbox{\scalerel*{
\begin{tikzpicture}[yscale=-1,transform shape]
\pic{orcidlogo};
\end{tikzpicture}
}{|}}}}

\usepackage[]{graphicx}    
\newcommand{\ignore}[1]{}  

\usepackage[xindy]{glossaries}
\loadglsentries{acronyms}
\usepackage{hyperref}
\usepackage{listings}
\usepackage{tabularx}
\usepackage{float}

\usepackage{doi}

\usepackage{fretish}

\usepackage[disable]{todonotes}

\begin{document}
\title{A Methodology for Developing a Verifiable Aircraft Engine Controller from Formal Requirements \thanks{
The authors would like to thank Georgios Giantamidis, Stylianos Basagiannis, and Vassilios A. Tsachouridis from United Technologies Research Center, Ireland for their participation in the requirements elicitation process for the aircraft engine controller use case.

This project has received funding from the ECSEL Joint Undertaking (JU) under grant agreement No 876852. 
The JU receives support from the European Union’s Horizon 2020 research and innovation programme and 
Austria, Czech Republic, Germany, Ireland, Italy, Portugal, Spain, Sweden, Turkey. Through ESCEL JU this project has received funding from Enterprise Ireland under grant agreement No IR20200054. Disclaimer: The ECSEL JU, the European Commission and Enterprise Ireland are not responsible for the content on this presentation or any use that may be made of the information it contains.}}

\author{%
Matt Luckcuck \orcidicon{0000-0002-6444-9312}\\
matt.luckcuck@mu.ie\\

\and
Marie Farrell \orcidicon{0000-0001-7708-3877} \\ 
marie.farrell@mu.ie\\

\and 
Ois\'in Sheridan\\
oisin.sheridan.2019@mumail.ie\\

\and
Rosemary Monahan \orcidicon{0000-0003-3886-4675 }\\
rosemary.monahan@mu.ie\\

~\\
Department of Computer Science, 
Maynooth University, Ireland
        
}

\date{\today}

\maketitle

\thispagestyle{plain}
\pagestyle{plain}


\begin{abstract}
\noindent Verification of complex, safety-critical systems is a significant challenge. Manual testing and simulations are often used, but are only capable of exploring a subset of the system's reachable states. Formal methods are mathematically-based techniques for the specification and development of software, which can provide proofs of properties and exhaustive checks over a system's state space. In this paper, we present a formal requirements-driven methodology, applied to a model of an aircraft engine controller that has been provided by our industrial partner. Our methodology begins by formalising the controller's natural-language requirements using the (pre-existing) \gls{fret}, iteratively, in consultation with our industry partner. Once formalised, \gls{fret} can automatically translate the requirements to enable their verification alongside a Simulink model of the aircraft engine controller; the requirements can also guide formal verification using other approaches. These two parallel streams in our methodology seek to combine the results from formal requirements elicitation, classical verification approaches, and runtime verification; to support the verification of aerospace systems modelled in Simulink, from the requirements phase through to execution. Our methodology harnesses the power of formal methods in a way that complements existing verification techniques, and supports the traceability of requirements throughout the verification process. This methodology streamlines the process of developing verifiable aircraft engine controllers, by ensuring that the requirements are formalised up-front and useable during development.
In this paper we give an overview of \gls{fret}, describe our methodology and work to-date on the formalisation and verification of the requirements, and outline future work using our methodology.
\end{abstract} 

\glsresetall


\section{Introduction}
\label{sec:intro}

The verification of complex systems presents a significant challenge to software engineering, particularly in safety-critical domains such as the aerospace sector. Industry often relies on manual testing and simulation to ensure that systems function correctly. However, these are especially time-consuming and expensive approaches that are only capable of exploring a subset of the system's reachable states.

Formal methods are a broad range of mathematically-based approaches for software engineering and verification. Historically, industrial uptake of formal verification approaches and methods has been slow. This may be due to the complexity, usability, or scalability issues sometimes faced by formal methods. However, formal methods are widely applicable to safety-critical applications in sectors like offshore oil and gas, the nuclear industry, and space exploration~\cite{luckcuck_formal_2019}\cite{fisher_overview_2021}\cite{luckcuck_using_2021}. They have also proven to be successful in many industrial projects~\cite{woodcock_formal_2009}.

Recent work argues for better integration of formal and non-formal verification
~\cite{farrell_robotics_2018}. One approach is \textit{corroborative verification}~\cite{webster_corroborative_2020}, where multiple verification techniques are applied to the system to each provide corroborative evidence that the system satisfies its requirements. This involves, for example, combining approaches that are less formal but are applied to concrete programs (such as simulation-based testing) with formal models, which often operate on more abstract representations of the system. Of course, this kind of mediation relies on a well-defined and unambiguous specification of the system's requirements, which has long been recognised as a bottleneck in the formal verification process~\cite{rozier_specification_2016}.

This paper contributes a three-phase methodology that connects a semi-formal requirements engineering phase with formal verification of a system's design. The link between these phases also encourages traceability of the requirements through to formalised properties that can be checked against the system and its design. We demonstrate our methodology through the verification of an aircraft engine controller, the requirements and designs of which have been supplied by our industrial partner.

Our methodology starts with the system's natural-language requirements, elicits further details, and encodes the requirements in a structured natural language.
This is a an iterative process that we have performed in close collaboration with our industrial partners, who work for United Technologies Research Center in Ireland. We use the \gls{fret}~\cite{giannakopoulou_formal_2020}, developed by a team at the NASA Ames Research Center in the USA 
for this initial phase.
Next, our methodology branches into two parallel phases. One phase uses requirements contracts that are automatically generated by \gls{fret} to check if a Simulink diagram of the system's design meets these requirements. The parallel phase uses the requirements expressed in \gls{fret} as the starting point for formal modelling of the system, so that other formal methods may be used to verify properties of the system. In this way we enable developers to adopt alternative verification approaches guided by the formalised \gls{fret} requirements. This methodology also supports a corroborative or integrated approach to verification allowing components of the system  to be verified using distinct techniques. Finally, we collect the verification artefacts and results produced to assemble a verification report. The nature of this report is for the developer to decide but we include it as a phase of our methodology for completeness.

Using a structured natural language with an underlying formal semantics to encode the requirements, rather than trying to formalise them directly, enables easier communication with our industry partners while still providing a language with enough structure to reduce ambiguity. This produces a clear set of requirements that are easier to formalise than the original natural-language requirements that they are based on, while retaining the collaborative nature of the requirements elicitation process. The two parallel phases in our workflow enable the system to be verified using a combination of complimentary approaches. Overall our methodology combines requirements elicitation, formal verification of Simulink designs, and classical formal methods.

Our methodology is inspired by previous work~\cite{bourbouh_integrating_2021}, which describes using \gls{fret} as the starting point for verification of Simulink designs and formal models. The contribution of this paper is to introduce a systematic methodology for driving the verification of a system's design from formalised requirements as well as facilitating heterogeneous verification.

\section{Background}

Formal Methods are a set of mathematically-based techniques for verifying that a software system behaves correctly. Typical examples of formal methods include theorem provers, which provide a proof of correctness of the system's behaviour; and model checkers, which exhaustively examine the system's state space to check that properties are preserved in all system executions. 

Dynamic properties modelling a systems safety are typically expressed in a temporal logic such as \gls{ltl} or \gls{ctl} \cite{fisher2011introduction}. These temporal logics specify changes in the state space of a system in response to events which occur \texttt{next}, \texttt{globally}, or \texttt{eventually}. Expressing safety properties in this way allows the use of theorem provers and model checkers to formally reason about the system's behaviour. Properties expressed in temporal logic will be generated as part of our methodology as we connect system requirements to the formal verification of that system's design.

Formal Methods are particularly beneficial in safety- and mission-critical domains such as aerospace, medical, transport, etc. because they provide a high degree of confidence in the correctness of the system. We aim to use these methods to verify a system's design, through a workflow that implements our methodology, using the tools introduced in this section.

\subsection{Writing Requirements in FRET}
\label{sec:fret}

\gls{fret} is an open-source tool that supports the writing of requirements in a structured natural language called \fretish{} \cite{giannakopoulou2020generation}. \gls{fret} helps us to bridge the gap between the English-language requirements provided by our industrial partner and the formal methods that we use during verification. In turn, this link improves the traceability of the requirements throughout the verification process.

\fretish{} is a structured English language, where requirements take the form: \\

\centerline{\fretishComponentsSmall{}}

The \Condition{}, \Component{}, and \Response{} parts of the requirement are mandatory; the keyword \texttt{shall} states that the component's behaviour must obey the requirement; and the other two parts of the requirement are optional. This allows requirements (described by the \Response{} clause) that are restricted to a certain \Scope{}, are triggered by a set of \Condition{}s, relate to a particular \Component{} of the system, may have particular \Timing{} constraints attached. 

\begin{figure*}
\center
\includegraphics[scale = 0.55]{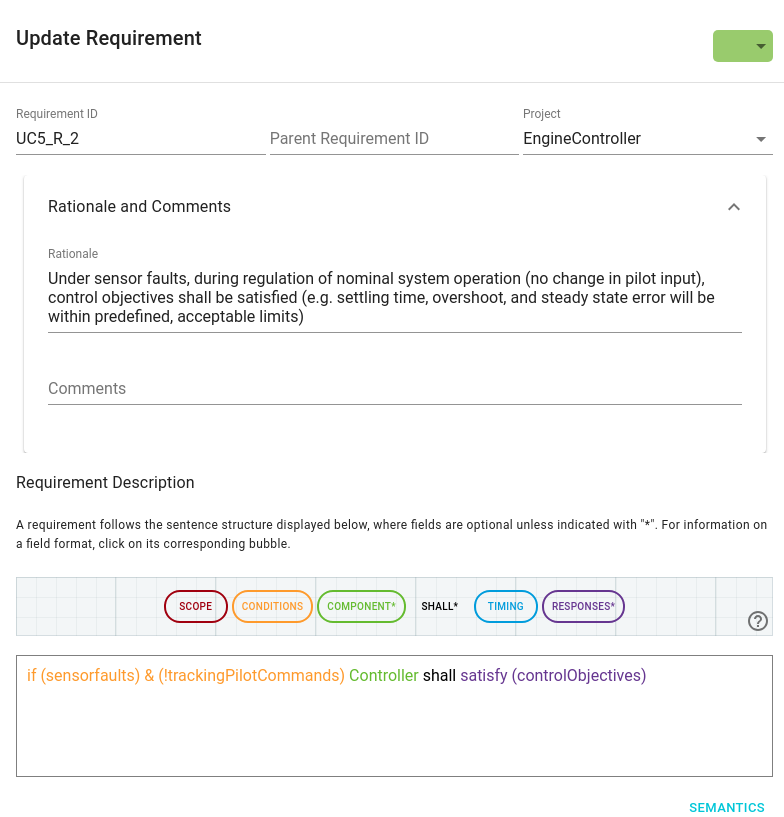}
\caption{\gls{fret}'s `Create Requirement' Dialogue \label{fig:fret-create} showing an example of a requirement in \gls{fret} with the requirement's ID, the  parent requirement ID, the project name, the requirement's description in \fretish{} as well as text boxes recording rationale and comments. 
}
\end{figure*}

Figure~\ref{fig:fret-create} shows the `Create Requirement' dialogue in \gls{fret} -- Figure~\ref{fig:fret-r1.1} shows the dialogue in more detail, which is explained in Sect.~\ref{sec:caseStudy}. It has fields for the requirement's ID, the ID of its parent requirement (if it has one), the project that it belongs to, and the requirement's description in \fretish{}. The `Rationale and Comments' item expands to allow the user to describe the requirement textually. This is useful for drafting notes and adding information that aids traceability.

For each requirement, \gls{fret} automatically generates formalisations in both past- and future-time metric \gls{ltl}. These formal translations mean that the requirements themselves can be reasoned about, formally, before the system's development progresses any further. The fact that this translation happens automatically speeds up this cycle of checks. \gls{fret} also displays a diagramatic semantics for each requirement. An example of this can be seen in Figure~\ref{fig:fret-r1.1}, which shows the time interval where the requirement should hold and the requirements triggering and stopping condition (if they exist). Both the temporal logic and diagramatic representations of the requirements are useful for understanding and sanity-checking the semantics of the requirement that has been written.


Our experience is that \gls{fret} and its in-built language, \fretish{}, provide a useful intermediate language to enable formalists and requirements engineers to work in collaboration towards formalising a system's requirements, without a steep learning curve for either team. The \fretish{} requirements provide a traceability link between natural-language and formalised requirements. \gls{fret} can generate output based on the requirements that is useful both directly and indirectly for verification. We describe the details of our workflow, beginning with using \gls{fret} in Sect.~\ref{sec:caseStudy}.

\subsection{CoCoSim and Simulink}

\gls{fret} integrates particularly closely with both Simulink and CoCoSpec~\cite{bourbouh2020cocosim}, the latter being a language for describing assume-guarantee contracts using Simulink blocks. \gls{fret} can be used to generate CoCoSpec contracts for each requirement, which can then be added to the Simulink diagram of the system that the requirements relate to. These contracts are checked during simulations of the diagram in Simulink using the Kind2 model checker \cite{champion2016kind}. In our methodology, presented in Sect.~\ref{sec:methodology}, we refer to this as the `\gls{fret}-Supported' toolchain.

Where Simulink is not used to design the system, or even where it is but other verification methods are desired, the temporal logic version of the requirements can be incorporated into other formal verification approaches. Where a formal verification tool directly accepts temporal logic, this will be straightforward. Otherwise, the temporal logic and \fretish{} versions of the requirements can be used as a guide for formalising the properties and system in other formalisms. In Sect.~\ref{sec:methodology} we refer to this as the `\gls{fret}-Guided' toolchain.

\section{Methodology}
\label{sec:caseStudy}

In this section we describe our methodology and our work to-date on the formalisation and verification of the requirements of our use case. First, we describe our use case, which is an aircraft engine controller. Then, we describe our methodology and how we have applied it to the requirements of the aircraft engine controller use case.

\subsection{Use Case: Aircraft Engine Controller}
\label{sec:controller}

Our example application is a software controller for a high-bypass civilian aircraft turbofan engine, which was provided by our industrial partner on the VALU3S project\footnote{The VALU3S project: \url{https://valu3s.eu/}}. The controller is a representative example of a Full Authority Digital Engine Control (FADEC) system, which monitors and controls everything about the engine, including thrust control, fuel control, power management, health monitoring of the engine, thrust reverser control, and so on. 

The controller's high-level objectives are that it should continue to control the engine (for example keeping settling time, overshoot, and steady state error within acceptable limits), respecting certain operating limits (for example respecting the engine shaft's upper speed limit), in the presence of:
\begin{itemize}
\item sensor faults (a sensor value deviating more than a given amount from its nominal value, or being unavailable),
\item perturbation of system parameters (a system parameter deviating more than a given amount from its nominal value), and
\item other low-probability hazards (for example, abrupt changes in the outside air pressure).
\end{itemize}
\noindent The controller is also required to detect faults and recover from them, which it does through a change of mode -- for example detecting engine surge or stall and changing modes to prevent these hazardous situations. The software controller supplied by our industrial partners is related to existing controllers previously described in the literature~\cite{postlethwaite1995digital}\cite{samar_design_2010}. 

\begin{table}
\begin{tabularx}{\textwidth}{|c|X|}
\hline
\textbf{ID} & \textbf{Description}\\ \hline
UC5\_R\_1 & Under sensor faults, while tracking pilot commands, control objectives shall be satisfied (e.g., settling time, 
overshoot, and steady state error will be within predefined, acceptable limits) \\ \hline
UC5\_R\_2 & Under sensor faults, during regulation of nominal system operation (no change in pilot input), control objectives 
shall be satisfied (e.g., settling time, overshoot, and steady state error will be within predefined, acceptable limits) \\ \hline
\end{tabularx}
\caption{The first two requirements for the aircraft engine controller, provided by our industrial partner. \label{tab:reqs} }
\end{table} 

Our industrial partner has supplied 14 English-language requirements and 20 test cases, which provide more detail about the controller's required behaviour. Table~\ref{tab:reqs} shows the first two requirements for the aircraft engine controller, provided by our industrial partner. As described in Sect.~\ref{sec:formalising}, in addition to the requirements and test cases, we iteratively formalised the requirements through elicitation meetings with our industrial partner. The controller's software has been designed in Simulink\footnote{Simulink: \texttt{\href{https://mathworks.com/products/simulink.html}{https://mathworks.com/products/simulink.html}}}. We focus on verifying that the high-level design for the controller obeys the requirements.

\subsection{Our Three-Phase Methodology}
\label{sec:methodology}

\begin{figure}
\centering
\includegraphics[width=0.47\textwidth]{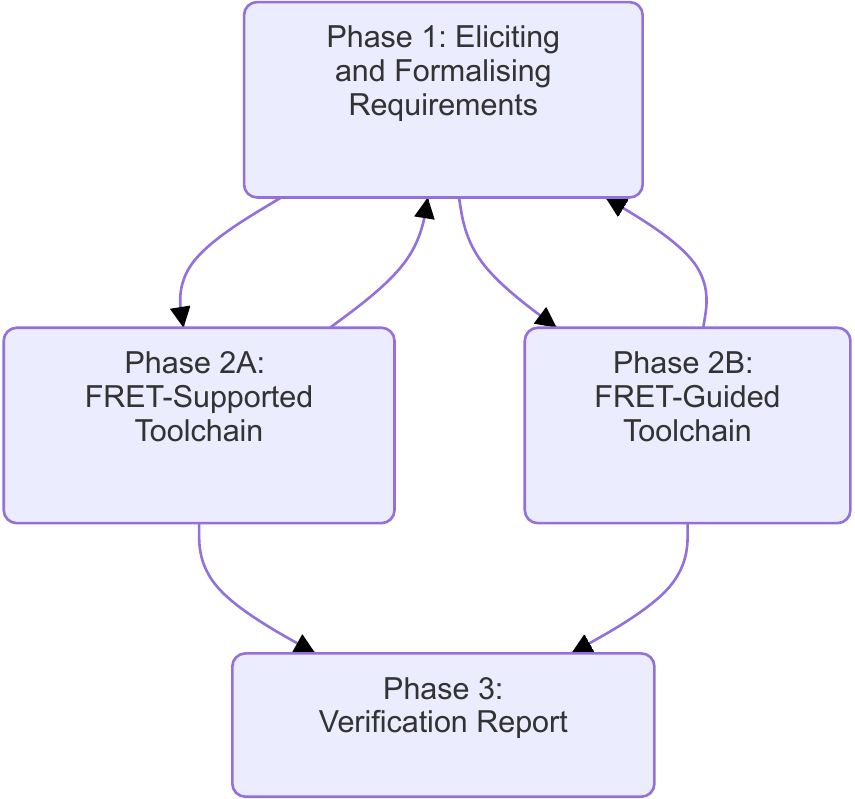}
\caption{High-Level Flowchart of our Methodology. After Phase 1 is complete, Phases 2A and 2B can occur in parallel. Phases 2A and 2B can both highlight deficiencies in the requirements, prompting a return to Phase 1. \label{fig:high-level-meth}}
\end{figure}

Our workflow takes requirements in natural-language and a Simulink diagram as input, and enables the formal verification of the system's design against the requirements. At a high level, our approach is split into three distinct phases, shown in Figure~\ref{fig:high-level-meth}. First we elicit and formalise the natural-language requirements using \gls{fret}, in Phase 1. Then we move on to formal verification either supported or guided by \gls{fret} (Phase 2A and 2B). The `\gls{fret}-Supported' toolchain  uses \gls{fret}'s built-in translation function to produce contracts that can be incorporated into a Simulink diagram. The `\gls{fret}-Guided' toolchain uses the formalised requirements to drive the (manual) translation into other formal methods as chosen by the verifier. Both verification phases can be applied in parallel. Finally, Phase 3 involves the assembly of a verification report to capture the verification results and traceability of requirements.

As outlined in Sect.~\ref{sec:intro}, our methodology `glues' the requirements engineering phase and formal verification phases together. These phases use different skills and would often be performed by different teams (or people). With this in mind, our methodology keeps the phases separate, but enables clear communication about the requirements and their formalisation using \fretish{} as a common language. This reduces the likelihood of incorrect or inaccurate requirements being formalised, while side-stepping the need for requirements engineers to learn a fully formal language.

In the remainder of this section we describe the workflow in detail. We make use of the aircraft engine controller, described earlier, as a running example.

\subsubsection{Phase 1: Eliciting and Formalising Requirements}
\label{sec:formalising}

Eliciting the system's requirements and formalising them is a task best completed by a collaboration between those knowledgeable about the system and those knowledgeable about formalisation. Using a formal language during this process can often slow down progress, because the systems engineers must learn the intricacies of the formal method and the formalists often try to formalise too much detail too quickly. Thus, in this phase we take the natural-language requirements as input, and produce the \fretish{} version of the requirements and the \gls{ltl} specifications of the requirements as output. Using \fretish{} as an intermediate language between the natural- and formal-language versions of the requirements, frees up the requirements elicitation process, but still provides enough formalisation in the requirements to both reduce ambiguities and make the requirements easier to formalise later on. 

\begin{figure}
\centering
\includegraphics[scale=0.5]{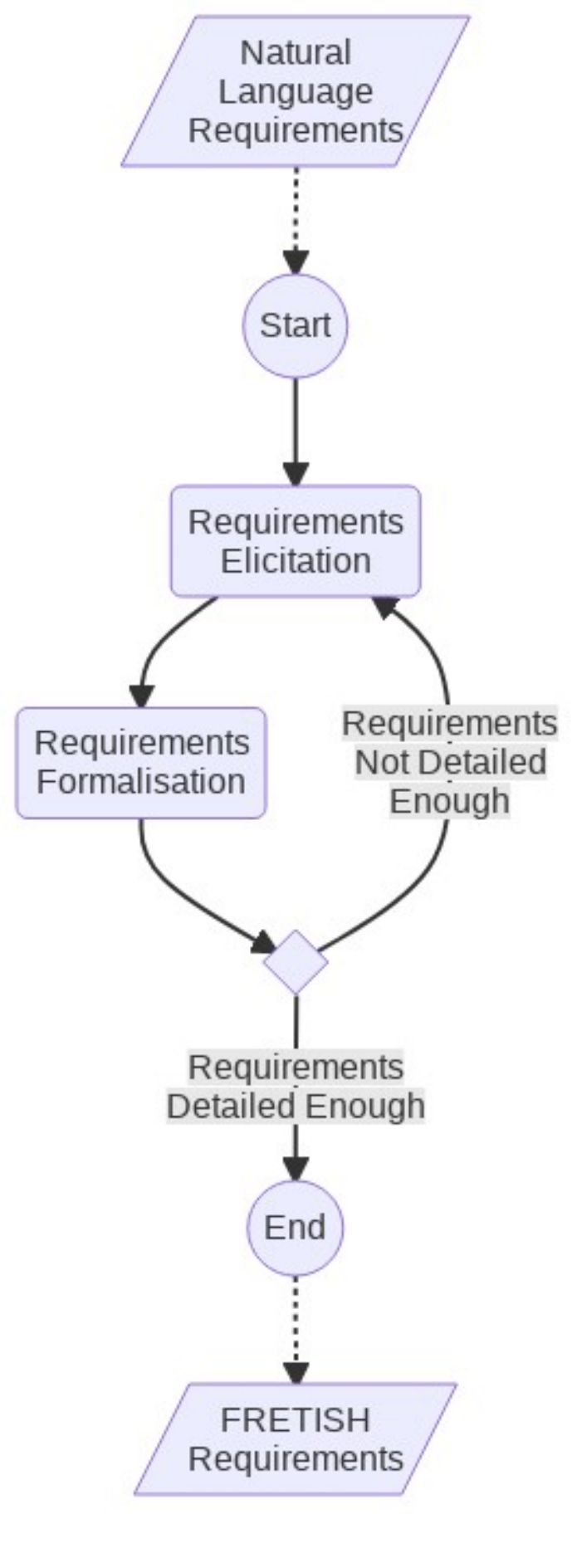}
\caption{Flowchart of Phase 1: Eliciting and Formalising Requirements. The circular nodes are start and end points, the rectangular nodes are processes (actions), and the rhomboid nodes are inputs or outputs. \label{fig:phase1} }
\end{figure}

The requirements elicitation phase will be slightly different for each system and set of requirements, but in our workflow (shown in Figure~\ref{fig:phase1}) we have separated this phase into a cycle of:

\begin{description}
\setlength{\leftskip}{0.3cm}
\item[\textsc{Step 1 (Requirements Elicitation):}] An informal process of discussing the natural-language requirements to clarify and reduce ambiguity, it also may uncover new requirements;\\

\item[\textsc{Step 2 (Requirements Formalisation):}] A manual process of converting the natural-language requirements into \fretish. 
\end{description}
\setlength{\leftskip}{0cm}
After Step 2, if the requirements have enough detail then the phase is over. If not, then the workflow cycles back to Step 1.

In our ongoing work, with our industrial partner, we have translated the 14 natural language requirements of the aircraft engine controller into \fretish. As an example of how this phase works, we will examine the first two requirements in our case study. Full details describing the requirements eliciation and formalisation process for this use case can be found in our report\footnote{\url{https://drive.google.com/drive/folders/1_5SUeh6B9jHrrEyKWN489-xA2zdCp6qx?usp=sharing}}. 

Requirement UC5\_R\_1 reads: ``Under sensor faults, while tracking pilot commands, control objectives shall be satisfied (e.g. settling time, overshoot, and steady state error will be within predefined, acceptable limits)'', which is expressed in \fretish{} as: 

\centerline{\small\texttt{\condition{((sensorfaults) \& (trackingPilotCommands))} \component{Controller} shall \response{(controlObjectives)}}}

Similarly, Requirement UC5\_R\_2 reads: ``Under sensor faults, during regulation of nominal system operation (no change in pilot input), control objectives shall be satisfied (e.g. settling time, overshoot, and steady state error will be within predefined, acceptable limits)", which is expressed in \fretish{} as: 

\centerline{\small\texttt{\condition{(sensorfaults) \& (!trackingPilotCommands)} \component{Controller} shall \response{(controlObjectives)}}}

\begin{figure*}
\center
\includegraphics[width=\textwidth]{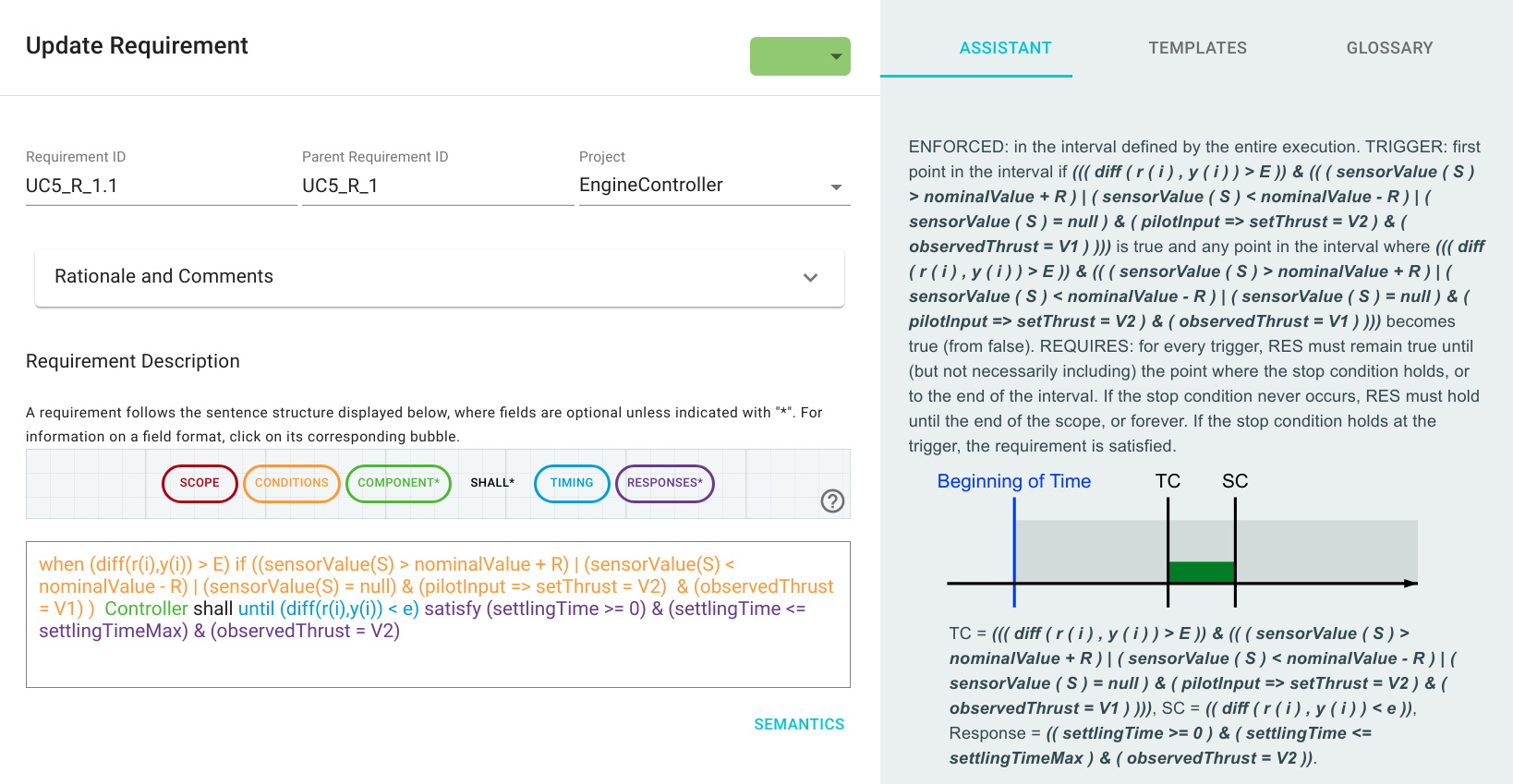}
\caption{The `Update Requirement' Dialogue Box in \gls{fret} for Requirement UC5\_R\_1.1, showing the requirement's ID, the  parent requirement ID, the project name, the requirement's description in \fretish{} as well as both the LTL representation of the requirements and the diagrammatic semantics for the requirement (on the right-hand panel) showing when the requirement becomes active and what its triggering (TC) and stopping (SC) conditions are.   
\label{fig:fret-r1.1}}
\end{figure*}

Through the iterations of requirements elicitation, we added more detail to the requirements, checking with our industrial partners that the requirements were correct. We added detail to the requirements through information gathered from the 20 test cases provided for the use case. For example, Test Case 1 (for Requirement UC5\_R\_1):

\begin{center}
\fbox{\parbox{0.8\textwidth}{
\begin{description}
\item[\textbf{Preconditions:}] Aircraft is in operating mode $M$ and sensor $S$ value deviates at most $\pm R \%$ from nominal value
\item[\textbf{Input conditions/steps:}] Observed aircraft thrust is at value $V1$ and pilot input changes from $A1$ to $A2$
\item[\textbf{Expected results:}] Observed aircraft thrust changes and settles to value $V2$, respecting control objectives (settling time, overshoot, steady state error)
\end{description}}
}
\end{center}

\noindent was split into three child requirements in \fretish. As an example, UC5\_R\_1.1 is the child requirement (of UC5\_R\_1) that adds details for the Control Objective \textit{settling time} is written as:

\begin{center}
{\texttt{\condition{when (diff(r(i),y(i)) > E) if ((sensorValue(S) > nominalValue + R) \\$|$ (sensorValue(S) < nominalValue - R) $|$ (sensorValue(S) = null) \\\& (pilotInput => setThrust = V2) \& (observedThrust = V1) ) }\\ \component{Controller} shall \timing{until (diff(r(i),y(i)) < e)} \\\response{(settlingTime >= 0) \& (settlingTime <= settlingTimeMax) \\\& (observedThrust = V2)}}}

\end{center}
Figure~\ref{fig:fret-r1.1} shows \gls{fret}'s `Update Requirement' dialogue for UC5\_R\_1.1. The ``Requirement Description'' section (bottom left) shows the \fretish{} version of the requirement that we described above. On the right-hand side, the semantics diagram (and accompanying textual description) displays when the requirement becomes active and what its triggering (TC) and stopping (SC) conditions are. 

Once the requirements elicitation and formalisation phase is complete, the requirements can be translated into the input language for the formal method(s) chosen for the verification phase.  Depending on the formal method chosen, this translation could be automatic or could be provided by the manual application of a set of systematic translation rules. 
If the design is described in a Simulink diagram, existing tool-support can take formalised requirements and produce contracts compatible with Simulink. Where the design has not been described using Simulink, or where the verification efforts would benefit from other formal approaches, we can build a formal model of the system from the available descriptions (the requirements, other models, documentation, etc.) and check the properties (formal encodings of the requirements) against the model.  

\subsubsection{Phase 2A: FRET-Supported Verification}
\label{sec:cocosim}

In this phase, we take the \fretish{} requirements and the Simulink design as input, and produce CoCoSpec contracts and verification results as output. The CoCoSpec contracts are built from Simulink blocks so that they can be embedded directly in a Simulink diagram. Because this phase is supported by \gls{fret}, much of the process is automated. However, the user still needs to provide information to link the \fretish{} requirements to the Simulink design, and the CoCoSpec contracts have to be manually added to the Simulink diagram for verification with the Kind2 model checker. 

\begin{figure}[t]
\centering
\includegraphics[scale=0.5]{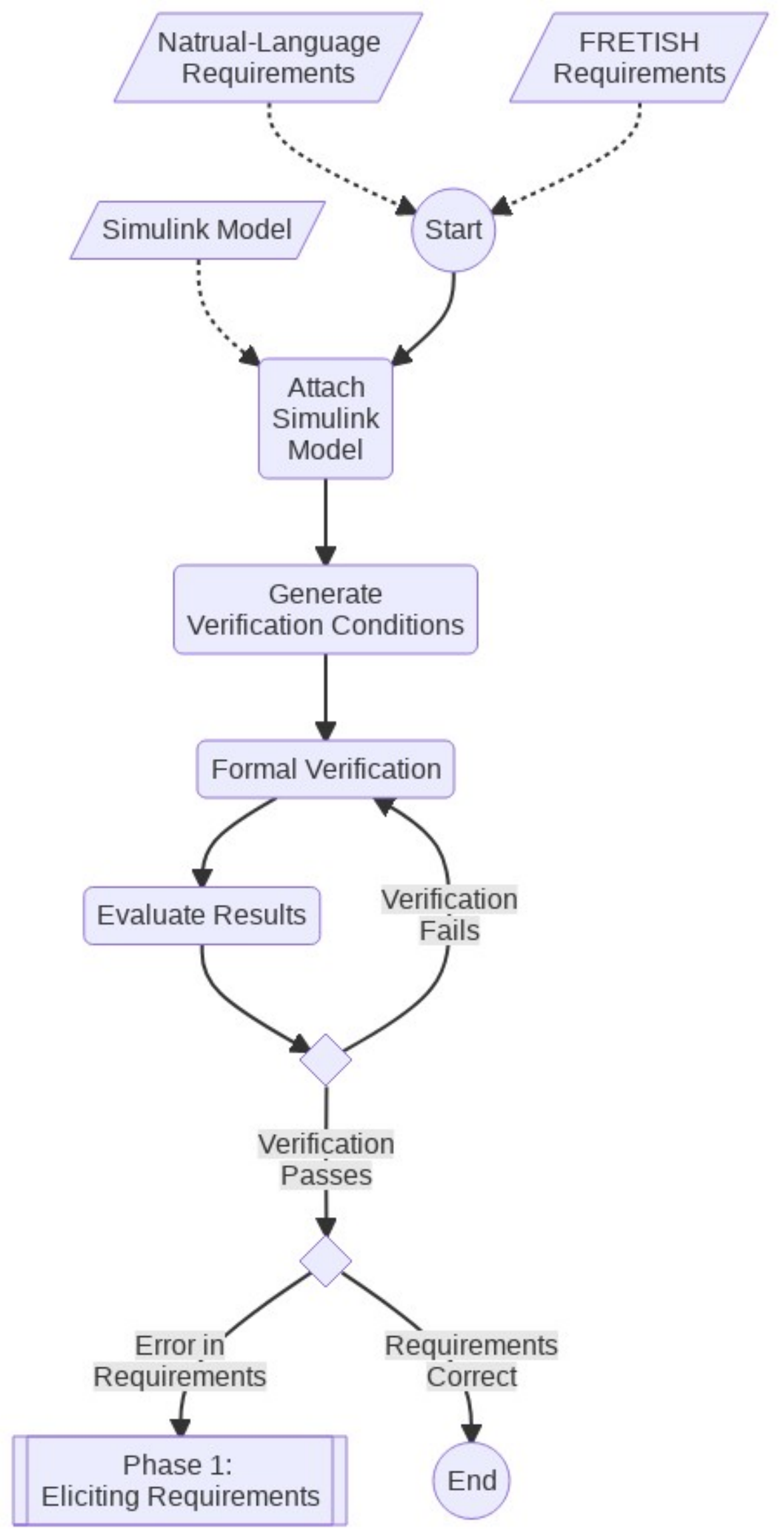}
\caption{Flowchart of Phase 2A: Verification supported by \gls{fret}'s link to CoCoSim. The circular nodes are start and end points, the rectangular nodes are processes (actions), the diamond nodes are decisions, and the rhomboid nodes are inputs or outputs. The double-edged rectangular node (bottom left) represents an external flowchart, in this case Phase 1. \label{fig:phase2a} }
\end{figure}

This phase of our workflow has four steps, which can be seen in Figure~\ref{fig:phase2a} and which we describe below.
\begin{description}
\setlength{\leftskip}{0.3cm}
\item[\textsc{Step 1 (Attach Simulink Model):}] \gls{fret} provides a Matlab script that extracts information from a Simulink diagram in a format that \gls{fret} can consume. Adding this to the \gls{fret} project makes it easier for the user to generate the verification conditions in the next step.\\

\item[\textsc{Step 2 (Generate Verification Conditions):}] In \gls{fret}, the user can add detail to link the requirements to the Simulink diagram of the system. For example, they can map a component mentioned in the requirements to a block or collection of blocks, or map a variable in the requirements to a signal in the Simulink diagram. Once this is complete, \gls{fret} can generate CoCoSpec contracts for the Simulink diagram, that represent the requirements. As an example, requirement UC5\_R\_1 (see Table \ref{tab:reqs}) is translated in the the CoCoSpec contract shown below:\\

{\small\texttt{guarantee "UC5\_R\_1" ((H(not(((sensorfaults) and (trackingPilotCommands))))) or \\(not(SI(((((sensorfaults) and (trackingPilotCommands))) and \\((pre(not(((sensorfaults) and (trackingPilotCommands))))) or FTP)), \\(not((controlObjectives)))))));}}\\

This CoCoSpec contract uses the operators: \texttt{H, SI, pre and FTP}. Here \texttt{H} represents the `Historically' past-time \gls{ltl} operator and \texttt{SI} represents the `Since Inclusive' operator. Then, \texttt{pre} is used to refer to the previous value of the variable following it and \texttt{FTP} refers to the First Time Point of the trace. The specific details of the CoCoSpec generation process are out of the scope of this work but we refer the interested reader to \cite{mavridou2020bridging} for the full details.\\

\item[\textsc{Step 3 (Formal Verification):}] Once the CoCoSpec contracts are incorporated into the Simulink diagram, they can be used to verify that the diagram (and so, the design of the system) obeys the contracts (and so, the requirements) during simulation runs using the Kind2 model checker.\\

\item[\textsc{Step 4 (Evaluate Results):}] If the verification fails, then it is possible that the Simulink diagram is incorrect, so it should be improved until the verification passes. It is also possible that the requirements are incorrect, in which case the workflow returns to Phase 1: Requirements Elicitation and Formalisation. If the verification passes, then the Simulink diagram obeys the requirements. 

\end{description}
\setlength{\leftskip}{0cm}

\subsubsection{Phase 2B: FRET-Guided Verification}
\label{sec:otherMethods}

In this phase we take both the natural-language and \fretish{} versions of the requirements as input, and produce formal properties, a formal model of the system, and verification results as output. This phase uses the \fretish{} requirements to drive formal modelling of the system.

\begin{figure}[t]
\center
\includegraphics[scale=0.5]{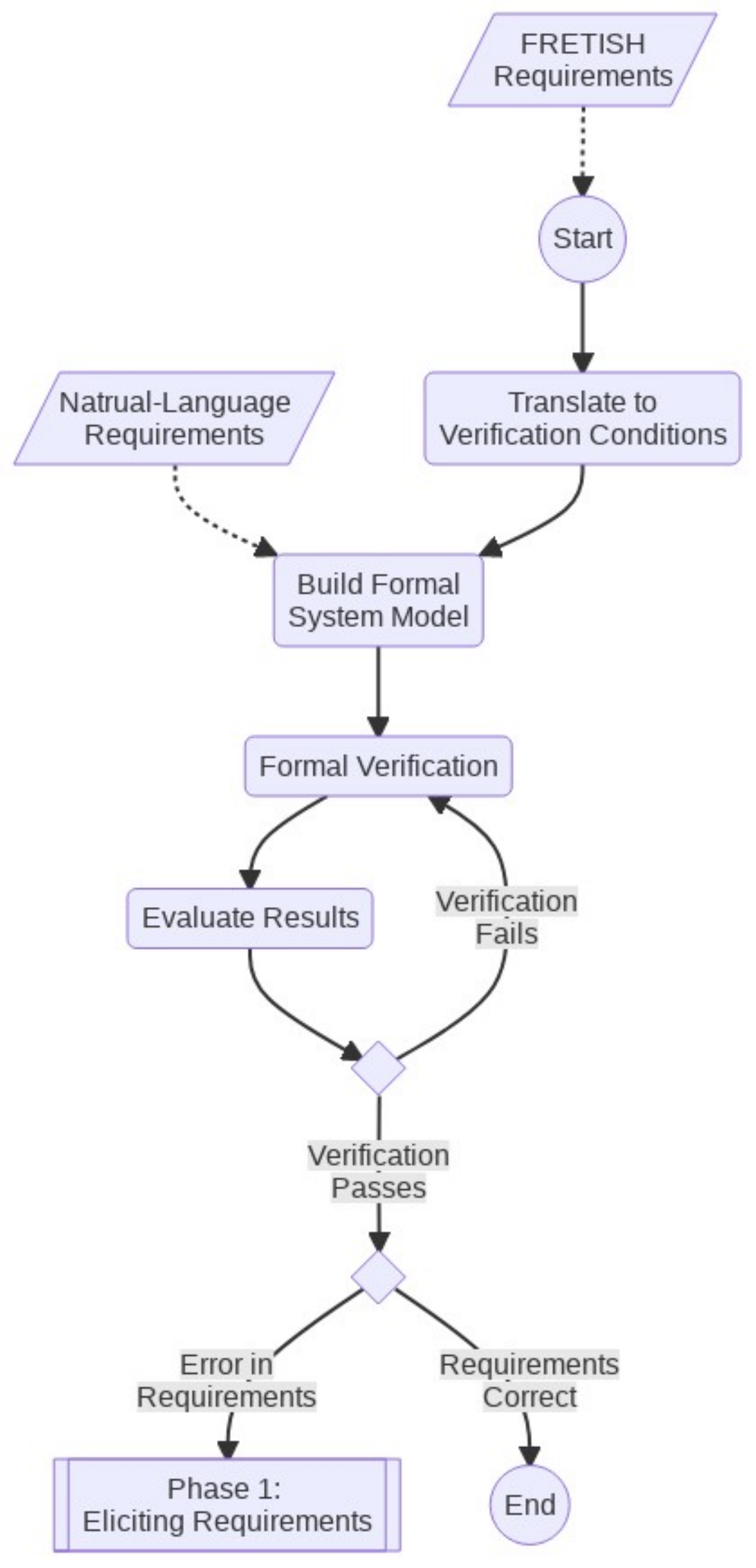}
\caption{Flowchart of Phase 2B: Verification guide by \fretish{} requirements. The circular nodes are start and end points, the rectangular nodes are processes (actions), the diamond nodes are decisions, and the rhomboid nodes are inputs or outputs. The double-edged rectangular node  (bottom left) represents an external flowchart, in this case Phase 1. \label{fig:phase2b} }
\end{figure}

We anticipate using formal methods that plug gaps in the verification approach used in Phase 2A. For example, methods that enable more detailed modelling of data, such as Event-B \cite{abrial2010modeling}; or more detailed modelling of behaviour, such as \gls{csp}~\cite{hoare1978communicating}. Again, this phase occurs in four steps, which can be seen in Figure~\ref{fig:phase2b} and we describe each of these individual phases below.

\begin{description}
\setlength{\leftskip}{0.3cm}
\item[\textsc{Step 1 (Translate to Verification Conditions):}] Using the \fretish{} requirements as the source, produce the properties to be verified in the chosen formal language. This is likely to be a non-formal translation process; but if the chosen formal method's input language is close to \fretish, then the translation will be straightforward. Obviously if the chosen formal method uses temporal logic, then the \gls{ltl} translations that \gls{fret} automatically produces could be used directly.\\

\item[\textsc{Step 2 (Build Formal System Model):}] Since this phase does not directly use a Simulink model, a formal model of the system or design must be constructed separately. This step is likely to involve using the \fretish{} and natural-language requirements, the Simulink model (if it exists), and other sources to ensure that the system model is valid.\\

\item[\textsc{Step 3 (Formal Verification):}] Using the appropriate tool (such as a model checker or theorem prover) perform the formal verification of the system model (Step 2) against the formal properties (Step 1).\\ 

\item[\textsc{Step 4 (Evaluate Results):}] As with the `\gls{fret}-Supported' toolchain in Phase 2A if the verification fails, then it is possible that the formal system model is incorrect and requires updates to make it pass the verification. It is also possible that the requirements themselves are wrong, and so the workflow returns to Phase 1: Requirements Elicitation and Formalisation. If the verification passes, then the system model obeys the formalised requirements. 
\end{description}
\setlength{\leftskip}{0cm}

\subsubsection{Phase 3: Verification Report}

Using the verification results returned by Phase 2A and 2B, along with the formalised requirements produced as part of Phase 1, a verification report is assembled. This report could take various forms, depending on what is required of the development process by laws, regulations, etc. Ensuring that verification (particularly formal verification~\cite{luckcuck_summary_2019}) can provide evidence of software's safety that is understandable by, and acceptable to, a regulator (or other organisation with supervisory responsibilities) is key in safety-critical domains. 

The report could be a natural-language report, structured as a safety case in a notation like the Goal Structuring Notation~\cite{kelly_goal_2004} or using tools such as the Assurance Case Automation Toolset (AdvoCATE) \cite{denney2012advocate}, or a combination of these formats. The key point is that the report facilitates traceability of the verified requirements and explains the extent of the verification activities, which may be useful during the assurance process. 

\section{Discussion}
\label{sec:discussion}

Our methodology, as presented in this paper, unifies and facilitates the use of multiple verification approaches whilst maintaining and supporting requirements traceability. We demonstrate our methodology through an industrial use case of an aircraft engine controller. Our methodology illustrates the benefits of using a formal approach to requirements capture so that formalised requirements can be directly fed into the verification process with minimal effort. 

Simulink is used extensively in the aerospace domain to model systems. We use FRET for requirements formalisation and elicitation because it has direct links to tools used for verifying Simulink diagrams. However, it has been shown that these tools are not always sufficiently scalable for complex requirements \cite{bourbouh_integrating_2021} so we also outline a pathway where other formal methods can be incorporated to support verification.

Our methodology provides a way to build and examine a series of \textit{prototype} systems/models before concrete implementation occurs. In particular, we can consider the \gls{fret} formalised requirements as a prototype formal model. They contain the conditions that the formal model must obey and some of the requirements even define how certain variables are updated/how the system should behave in a given context. Then, we can analyse the Simulink model which is a prototype implementation that can be examined with respect to the generated CoCoSpec contracts. In addition, we can build formal models of the system which can be treated as prototype implementations since they mirror the behaviour of the final system, with some details abstracted. 

\subsection{Evolution of Requirements}

Our high-level methodology, as captured in Figure~\ref{fig:high-level-meth}, contains optional loops that go back from the verification phases (Phases 2A and 2B) to the requirements elicitation phase (Phase 1). The option to return to this earlier phase enables the methodology to respond to errors that are only identified in Phase 2A or 2B, but indicate incorrect or inaccurate requirements. It is important that the requirements are captured as accurately as possible; if errors or omissions in the requirements are found, then any changes to remedy these problems may also propagate to the accompanying Simulink diagram(s) or formal model(s). However, such changes would also propagate if a non-formal requirements engineering approach were being used.

\gls{fret} provides some assistance with changes propagating from updated requirements to their derived models and diagrams.
First, when a \fretish{} requirement is updated, \gls{fret} automatically converts it to \gls{ltl}, so there is no user overhead in this step. The updated requirement can be re-exported to CoCoSpec (Phase 2A), or the updated requirement can be re-translated into the formal language used before (Phase 2B). These are manual steps, but since the contracts or formal properties have been built from each of the \fretish{} requirements, the traceability provided by \gls{fret} makes their regeneration more straightforward than if they had been built directly from the natural language requirements.

In addition to revision of the verification conditions, revision of the system's design and associated models may be needed if the requirements are found to be wrong. Again, this would be a manual process of updating the design to match the changes to the requirements, but this would also be required in a non-formal approach. In practice, software development follows an iterative process where additional requirements may be added or existing ones modified as development continues and errors or inconsistencies are revealed in the design. 
Here, \gls{fret} provides traceability features that make these updates easier to implement. Requirement IDs enable links between requirements to be traced, the source of the requirement (for example its natural-language counterpart) can be captured in the rationale field, and the CoCoSpec contracts that \gls{fret} automatically produces share the name of the requirement from which they were generated. These features mean that the \fretish{} requirements can be linked back to the natural-language requirements, and linked forward to the verification conditions and system designs, therefore finding the target for updates is relatively straight-forward.

\subsection{\gls{fret}-Supported vs \gls{fret}-Guided Verification}
\label{sec:OneLegOrTwo}

As previously mentioned, Phase 2A (\gls{fret}-Supported Verification) and Phase 2B (\gls{fret}-Guided Verification) can be applied separately or together in parallel. Both phases apply formal verification to the system's design, but using different languages to specify the properties and using different ways of describing the design itself. Here we discuss the interactions between these two phases, the situations where each might be useful, and the benefits of applying both phases to a system's design. 

The main difference between the two verification phases are in the targets of the verification. Phase 2A targets a Simulink diagram, using the Kind2 model checker. In contrast, Phase 2B targets a separate formal model of the system, built in a formal method of choice. Direct use of a Simulink diagram as the description of the system is useful (if the diagram exists), but the available checks are limited to those directly describable in \fretish{}. Introducing another formal language, and its associated tools, might be advantageous. Using the approach supported in Phase 2B may enable the verification of properties that it was not possible to verify using the approach supported in Phase 2A, as was found to be the case in related work~\cite{bourbouh_integrating_2021}.

Using both phases in parallel provides a more corroborative approach to verification, where different styles of verification are used to corroborate each other~\cite{webster_corroborative_2020}. Corroborative verification is often focussed on applying multiple verification techniques to the same set of properties, but applying both phases in our methodology also offers the developer the chance to pick and choose which techniques to use to verify which specific requirements. In this `divide and conquer' strategy, a formal method can be chosen that complements the CoCoSpec approach and plugs any gaps or limitations that it may have, as was the approach in \cite{bourbouh_integrating_2021}.

Further, using both phases produces two independent designs, both derived from the same set of formalised requirements. Comparing the design in Simulink to a design in a formal language may help to identify inconsistencies or ambiguities in either design, or in the requirements themself, that were not identified during requirements elicitation. Exploring different designs in this way may produce more robust implementations, because the requirements have been viewed through a variety of lenses.






\section{Related Work}
\label{sec:related}

In this section we describe some of the related work in the literature, focussing on the themes of requirement specification, the formal verification of aircraft systems, and the formal verification of Simulink diagrams.

\subsection{Requirements Specification}
Specifying verifiable and precise requirements can often delay the verification process \cite{rozier_specification_2016}. Requirements elicitation and formalisation tools can aid this process by helping the user to define unambiguous requirements from the outset. In safety- or mission-critical domains such as the aerospace sector, development is often driven by high-level requirements that are gradually decomposed as the project continues.

Some approaches to requirements engineering simply use natural-language. For example, IBM's Dynamic Object-Oriented Requirements System (DOORS)\footnote{\url{https://www.ibm.com/docs/en/ermd/9.7.0?topic=overview-doors}} is frequently used in industry to manage and support traceability of requirements. In DOORS, requirements are represented in natural-language without any view to formalise them or reduce ambiguities so that they can be easily used by formal methods.

In contrast, there have been multiple approaches developed that support the use of more formal, logical requirements. These include \gls{fret}, which uses a structured natural language to provide an \gls{ltl} semantics for formalised requirements \cite{giannakopoulou2020generation}. \gls{fret} uses the Structured Assertion Language for Temporal Logic (SALT), which is a general purpose assertion language with property pattern features as an intermediate language \cite{bauer2006salt}.

Formalised requirements have also been added to existing, non-formal approaches. For example, the Easy Approach to Requirements Syntax (EARS) uses informal templates to define high-level requirements \cite{mavin2012listen}. However, recent work has provided a formalisation for these templates in \gls{ltl} \cite{lucio2017just}. Despite this advancement, EARS does not support contract generation whereas \gls{fret} does.


\subsection*{Formal Verification of Aircraft Systems}

Formal verification tools and techniques have been applied to an array aircraft systems. This inlcudes the specification and verification of a landing gear system in Event-B \cite{mammar2017modeling} and Abstract State Machines \cite{kossak2014}. Related work focuses on the verification of a sensor voting module used in the landing gear system using model-checking with Asmeta \cite{Arcaini_2014}.

An \gls{aadl} Model of an F-16 Autopilot Controller has been translated into Stateful Timed \gls{csp}, which was then model checked in the Process Analysis Toolkit (PAT) checker. The Autopilot Controller is a non-linear model that simulates the dynamics of the real aircraft. This work focuses on verifying safety, liveness, and trace refinement properties \cite{Zhang_2017}. Other work that incorporates \gls{csp} includes the use of Hybrid \gls{csp}  to model and verify an aircraft system. The example system is a controller for correcting an aircraft's position when it has deviated from its intended flight path \cite{Peng_2015}. 

In \cite{backes2015requirements}, Backes et al. use the  Assume Guarantee REasoning Environment (AGREE) \cite{cofer2012compositional} for compositional verification of an \gls{aadl} model of a Quad-Redundant Flight Control System. They use the Kind2 and JKind model-checkers for verification. One of the main difficulties that they had was in formalising natural language requirements at a level of abstraction and detail that was amenable to compositional verification. Our work seeks to streamline this process by including detailed requirements formalisation from the outset. 

\subsection*{Formal Verification of Simulink Diagrams}

The popularity of Simulink for designing systems has lead to a variety of formal verification techniques that target Simulink diagrams. In addition to its own Simulink Design Verifier \cite{hamon2008simulink}, the literature contains examples both of model checking and theorem proving as the formal verification approach.

Some of the approaches in the literature translate the elements of a Simulink diagram into the input language of a theorem prover. For example, Herber et al. 2013~\cite{herber_bit-precise_2013} present an automatic translation from discrete-time Simulink diagrams to the input language of the UCLID system verification toolkit\footnote{UCLID: \url{https://github.com/uclid-org/uclid}}. Similarly, Reicherdt and Glensner 2014~\cite{reicherdt_formal_2014} translate discrete-time Simulink diagrams into Boogie, an intermediate language for the Z3 verification framework\footnote{Boogie: \url{https://www.microsoft.com/en-us/research/project/boogie-an-intermediate-verification-language/?from=http\%3A\%2F\%2Fresearch.microsoft.com\%2Fen-us\%2Fprojects\%2Fboogie\%2F}}.

Meenakshi et al.~\cite{meenakshi_tool_2006} present a tool that automatically translates discrete-time Simulink diagrams into the input language of the model checker NuSMV\footnote{NuSMV: \url{https://nusmv.fbk.eu/index.html}}. Their tool parses information about the Simulink diagram from its textual representation in an \texttt{mdl} file, then translates it for NuSMV. The restriction to discrete-time diagrams is due to NuSMV only catering for discrete models. Their tool can also reverse the translation, to aid with debugging the Simulink diagram from counterexamples. 

To capture the continuous-time elements of Simulink diagrams, Liebrenz et al. 2018~\cite{liebrenz_deductive_2018} provide an automatic translation to \gls{dl} to enable interactive theorem proving using KeYmaera X\footnote{KeYMaera X: \url{https://www.keymaerax.org/index.html} }. However, their automatic translation assumes: no algebraic loops, no S Function blocks\footnote{S (System) Function blocks wrap a program, written in either: MATLAB, C C++, or FORTRAN code}, and no external scripts or libraries. 

Instead of translating Simulink diagrams into another language, Bernardeschi et al. 2018~\cite{bernardeschi_pvs-simulink_2018} use co-simulation, where Simulink diagrams are simulated in parallel with a model in the PVS theorem prover\footnote{PVS: \url{http://pvs.csl.sri.com/}.}. The PVS model captures the discrete semantics of the system and the Simulink model captures the continuous dynamics of the system's environment. They use PVSio-Web\footnote{PVSio-Web: \url{http://www.pvsioweb.org/}}, which is a Web framework that enables formal-model-based development of Human-Machine Interfaces, to connect the PVS model to the Simulink diagram. They also use an additional block in the Simulink diagram to extract the simulation data. PVSio-Web uses this block's output to keep the PVS model in-sync with the Simulink simulation.

The addition of blocks to the Simulink diagram is the approach taken by CoCoSpec~\cite{bourbouh2020cocosim}, which we use in our methodology. This approach has already been described in the literature, for example In Araiza-Illan et al. 2014~\cite{araiza-illan_formal_2014} where assertion blocks are added to Simulink diagrams to describe the properties (over the Simulink signals) that the system should preserve. These assertions take the form of a \texttt{require} block, which is used to describe a Hoare triple: $\{precondition\} model \{postondition\}$. A \texttt{require} block is constructed from an \texttt{Enabled Subsystem}, which conditionally executes when a control signal (the precondition) is positive and itself contains an \texttt{Assert} block. Once this specification is added to the Simulink diagram, they provide an automatic translation into the input language of the Why3\footnote{Why3: \url{http://why3.lri.fr/}} theorem proving platform, and used CVC3\footnote{CVC3: \url{https://cs.nyu.edu/acsys/cvc3/}} to verify the Simulink diagram against the properties described by the assertion blocks. Their translation strategy works on Simulink signals that are scalar and discrete, but they aimed to extend the approach to continuous time. 


\section{Conclusions and Future Work}
\label{sec:conclusion}
This paper has presented our methodology for developing a verifiable system using a formal requirements-driven approach. This work is targetted for the aerospace domain and we illustrate our methodology via an aircraft engine controller use case that has been supplied by our industry partner, United Technologies Research Center in Ireland. 

Our methodology is rooted in requirements elicitation and formalisation, to drive the derivation of appropriate verification conditions and support traceability. We have chosen \gls{fret} as the basis for this approach because it has support for verifying Simulink diagrams and previous work has shown that \gls{fret} requirements are relatively straightforward to incorporate into other verification formalisms \cite{bourbouh_integrating_2021}. In saying that, future work will seek to make this link between \gls{fret} and other formal methods more seamless.

Our methodology provides an avenue to harness the power of formal methods in a way that complements existing verification techniques, which streamlines the process of developing verifiable aircraft engine controllers. Further, the requirements-driven approach that we adopt supports the traceability of requirements throughout the verification process. The combination of these existing tools applied to an industrial aircraft engine controller has, to the best of our knowledge, not been explored before. We also believe that our approach is flexible and, although we focus on a particular suite of tools, that other tools may be easily incorporated where necessary.

We are continuing to elaborate this use case alongside our industry partner whose primary focus is the aerospace domain. However, we also intend to investigate other domains which make extensive use of Simulink and related approaches to examine and evaluate our work more broadly.






\bibliographystyle{plainurl}
\bibliography{verification-methodology}

\end{document}

%% file: acronyms.tex



\newacronym{ltl}{LTL}{Linear-time Temporal Logic}

\newacronym{ctl}{CTL}{Computation Tree Logic}

\newacronym{ptl}{PTL}{Probabilistic Temporal Logic}

\newacronym{pctl}{PCTL}{Probabilistic Computation Tree Logic}

\newacronym{pctl*}{PCTL*}{Probabilistic Computation Tree Logic*}

\newacronym{tctl}{TCTL}{Timed Computation Tree Logic}

\newacronym{stl}{STL}{Signal Temporal Logic}

\newacronym{ltl-x}{LTL-X}{a version of Linear Time Logic without the `next' operator}

\newacronym{mtl}{MTL}{Metric Temporal Logic}

\newacronym{iactlk-x}{IACTLK-X}{a fragment of the temporal-epistemic logic CTLK, without the `next' operator}

\newacronym{mucalc}{$\mu$-calculus}{a strict superset of other temporal logics}

\newacronym{fotl}{FOTL}{First-Order Temporal Logic}

\newacronym{bdi}{BDI}{Belief-Desire-Intention}

\newacronym{prs}{PRS}{Procedural Reasoning System}

\newacronym{are}{ARE}{Autonomous Reasoning Engine}

\newacronym{dmars}{dMARS}{distributed Multi-Agent Reasoning System}


\newacronym{fsm}{FSM}{Finite-State Machine}

\newacronym{pfsm}{PFSM}{Probabilistic Finite-State Machine}

\newacronym{apfsm}{APFSM}{\textit{Autonomous} Probabilistic Finite-State Machine}

\newacronym{pha}{PHA}{Probabilistic Hybrid Automata}

\newacronym{fsa}{FSA}{Finite-State Automata}

\newacronym{ta}{TA}{Timed Automata}

\newacronym{gspn}{GSPN}{Generalised Stochastic Petri Net}

\newacronym{mopn}{MOPN}{Marked Ordinary Petri Net}

\newacronym{ba}{BA}{B\"{u}chi Automata}

\newacronym{dtmc}{DTMC}{Discrete-Time Markov Chain}


\newacronym{pars}{PARS}{Process Algebra for Robot Schemas}

\newacronym{fsp}{FSP}{Finite State Processes}

\newacronym{piadl}{$\pi$ADL}{$\pi$ calculus combined with ADL}

\newacronym{csp}{CSP}{Communicating Sequential Processes}

\newacronym{ccs}{CCS}{Calculus of Communicating Systems}

\newacronym{wsccs}{WSCCS}{Weighted Synchronous CCS}

\newacronym{csp|b}{CSP$\|$B}{an integrated formalism combining CSP and B}


\newacronym{fdr}{FDR}{the Failures-Divergences Refinement checker}

\newacronym{jpf}{JPF}{Java PathFinder}

\newacronym{ajpf}{AJPF}{Agent Java PathFinder}

\newacronym{mcmas}{MCMAS}{Model Checker for Multi-Agent Systems}

\newacronym{ltlmop}{LTLMoP}{Linear Temporal Logic MissiOn Planning}


\newacronym{adl}{ADL}{Architecture Description Language}

\newacronym{aadl}{AADL}{Architecture Analysis and Design Language}

\newacronym{lts}{LTS}{Labelled-Transition System}

\newacronym{uml}{UML}{Unified Modelling Language}

\newacronym{sysml}{SySML}{Systems Modelling Language}

\newacronym{robotml}{RobotML}{Robot Modelling Language}

\newacronym{dsl}{DSL}{Domain Specific Language}

\newacronym{sct}{SCT}{Supervisory Control Theory}

\newacronym{ide}{IDE}{Integrated Development Environment}

\newacronym{mde}{MDE}{Model-Driven Engineering}

\newacronym{ai}{AI}{Artificial Intelligence}

\newacronym{fdir}{FDIR}{Fault Detection, Isolation and Recovery}

\newacronym{ifm}{iFM}{Integrated Formal Methods}

\newacronym{ros}{ROS}{Robot Operating System}

\newacronym{amcl}{AMCL}{Adaptive Monte Carlo Localisation}

\newacronym{dl}{\text{\upshape\textsf{d{\kern-0.05em}L}}}{differential dynamic logic}

\newacronym{vm}{VM}{Virtual Machine}

\newacronym{bip}{BIP}{Behaviour Interaction Priority}

\newacronym{fol}{FOL}{First-Order Logic}

\newacronym{owl}{OWL}{Web Ontology Language}

\newacronym{ria}{RIA}{Restore Invariant Approach}

\newacronym{ocl}{OCL}{Object Constraint Language}

\newacronym{sil}{SIL}{Safety Integrity Level}

\newacronym{rcl}{RCL}{ROS Contract Language}

\newacronym{rv}{RV}{Runtime Verification}

\newacronym{rml}{RML}{Runtime Monitoring Language}